\documentclass[12pt]{iopart}
\usepackage{amssymb}
\usepackage{graphicx}
\usepackage{hyperref}

\newcommand{\be}{\begin{equation}}
\newcommand{\ee}{\end{equation}}
\newcommand{\aky}{k_y \rho_i}
\newcommand{\altic}{a/L_{T_i,\mathrm{critical}}}
\newcommand{\alti}{a/L_{T_i}}
\newcommand{\alte}{a/L_{T_e}}
\newcommand{\altz}{a/L_{T_z}}
\newcommand{\altj}{a/L_{T_j}}
\newcommand{\alni}{a/L_{n_i}}
\newcommand{\alne}{a/L_{n_e}}
\newcommand{\alnz}{a/L_{n_z}}
\newcommand{\alnj}{a/L_{n_j}}
\newcommand{\nzne}{n_z/n_e}
\newcommand{\vthi}{v_{\mathrm{th},i}}

\newcommand{\chiGB}{\chi_\mathrm{GB}}

\begin{document}

\title{On the mutual effect of ion temperature gradient instabilities and impurity peaking in the reversed field pinch}

\author{I. Predebon$^1$, L. Carraro$^1$, C. Angioni $^2$}
\address{$^1$ Consorzio RFX, Associazione EURATOM-ENEA sulla fusione, Padova, Italy}
\address{$^2$ Max Planck Institut f\"ur Plasmaphysik, EURATOM Association, Boltzmannstrasse 2, 85748 Garching, Germany}

\date{\today}

\begin{abstract}
The presence of impurities is considered in gyrokinetic calculations of ion temperature gradient (ITG) instabilities and turbulence in the reversed field pinch device RFX-mod. This device usually exhibits hollow Carbon/Oxygen profiles, peaked in the outer core region. We describe the role of the impurities in ITG mode destabilization, and analyze whether ITG turbulence is compatible with their experimental gradients.
\end{abstract}


\maketitle


\section{Introduction}
Impurity transport in magnetic fusion devices is a long-time debated subject, which in diverted tokamaks gains topical relevance to avoid high effective charge accumulation in the plasma core. This has been a rich field of research in experimental fusion science for many years, with remarkable attention to the role of auxiliary heating systems on the impurity flux (often reversing impurity convection from inwards to outwards)~\cite{rice02, dux03, takenaga03, puiattivalisa06}, and to the role of edge~\cite{belo04} and internal~\cite{dux04} transport barriers in inducing impurity flattening or accumulation. Much of the related theoretical activity has concentrated on the role of neoclassical transport~\cite{fussmann}, and more recently on that of electrostatic ITG/TEM turbulence~\cite{angioniprl06, camenen09, futatani10, valisa11}, but often with contrasting results. Viceversa, whether large impurity concentration may yield turbulence enhancement or quenching, is an even more open issue. In the context of electrostatic turbulence the reversed field pinch (RFP) is a scarcely explored configuration -- transport being historically thought as dominated by large-scale MHD turbulence -- which may give a contribution to the knowledge of impurity transport in toroidal fusion plasmas.

Recently the study of microturbulence has in fact extended to the RFP configuration. The main reason is the achievement of core plasmas naturally endowed with almost conserved magnetic flux surfaces. In the RFX-mod device~\cite{martin09}, for example, this has been accomplished thanks to the self-organized single helicity states, in which a single $m=1$ kink-tearing mode is able to dynamically sustain the discharge. The net effect is the arising of an overall helical warping, usually with the periodicity of the innermost $m=1$ resonant mode. When the plasma current is high enough, the plasma has a further transition towards the so called single helical axis state~\cite{lorenziniNature}, with separatrix expelled and a rather peculiar electron temperature profile, flat in the core and barrier-like at mid radius. Present diagnostics at RFX allow for determining the electron temperature only, but measurements from passive spectroscopy let estimate for the electron to ion temperature ratio $T_e/T_i$ a value about 3/2 in the core.

Does microturbulence really play an important role in RFP transport? Previous studies have shown that ion temperature gradient (ITG) instabilities~\cite{romanelli89, horton99} can be considered only a partial contributor to particle/energy transport in the RFP~\cite{guo08, predebon10a, sattin10a, sattin10b}, microtearing modes being much more relevant~\cite{predebon10b}. This has been explained as a consequence of the significant Landau damping of the ITG mode, driven by the field connection length $\sim 2\pi a$ ($a$ is the torus minor radius), or equivalently by the low value of the safety factor, $q\ll 1$. Part of this paper is devoted to a more careful analysis of the ITG stability threshold. In particular we will investigate the role of the impurities in ITG mode destabilization. Such a destabilization may occur in RFX-mod, due to the hollow profile of the impurities~\cite{liu11} caused by their inefficiency in reaching the inner plasma~\cite{menmuir10}. For this analysis we adopt the code GS2~\cite{krt, dorland00}, a widely known gyrokinetic code based on the electromagnetic nonlinear gyrokinetic equation~\cite{antonsenlane, friemanchen, brizardhahm}, solved in a flux tube domain. The modifications required for GS2 to work in the RFP configuration are examined in Ref.~\cite{predebon10a}, and mainly concern the large poloidal component of the magnetic field. The RFX-mod single helicity fields have been approximated as axisymmetric: 3D gyrokinetic simulations based on reversed field pinch VMEC equilibria~\cite{terranova10, gobbin11} is an upgrade to look at for future studies.

In RFX-mod the most populated impurity density profiles are hollow, peaked in the outer core domain $0.6\lesssim r/a \lesssim 0.9$. This feature is encountered in all the RFX-mod experimental regimes, either during the chaotic multiple-helicity stages or in the laminar quasi single helicity states~\cite{menmuir10}. The absence of stationary impurity accumulation is an appealing feature for any fusion device, which makes this RFP machine an interesting test bed both for experimentalists and theoreticians. Since neoclassical transport is practically negligible in this configuration, impurities turn out to be driven either by some magnetic chaos in the outer core region or, possibly, by small scale micro-turbulence, an option never explored so far. In this work we are going to look into the latter possibility, neglecting the effects related to magnetic surface destruction due to tearing or microtearing instabilities. In particular we will consider the role played by purely electrostatic ITG turbulence. Dealing only with this kind of turbulence can be heuristically considered relevant if the resulting ion thermal diffusivity is much larger than the cross-field ion diffusivity due to magnetic chaos. Assuming a gyro-Bohm scaling for the former ($\sim\rho_i^2\vthi/L_{T_i}$, with $\rho_i$ ion Larmor radius, $\vthi$ ion thermal speed, $L_{T_i}$ ion temperature gradient scale length) and a collisionless quasi-linear expression for the latter ($\sim b^2 L_c \vthi$, with $b$ normalized magnetic field perturbation and $L_c$ parallel correlation length), such condition roughly reads $(\rho_i/a)^2 (\alti)/b^2 \gg 1$, where we have considered $L_c\sim a$, a valid estimate for both tearing and microtearing turbulence in RFX-mod. An alternative to get the same condition is to require the ITG mode growth rate to be much larger than the rate needed for the ions to spread radially by $\sim 1-10\,\rho_i$ ($\sim k^{-1}_{\perp\;\mathrm{ITG}}$) due to magnetic diffusion. The inequality above is easily fulfilled in the case of micro-tearing turbulence, where $b\sim \rho_e/L_{T_e}$; conversely, for a large scale tearing turbulence $b$ is much larger, often yielding $(\rho_i/a)^2 (\alti)/b^2 \lesssim 1$. Thus, as regards ion transport, neglecting magnetic macro-turbulence turns out to be the strongest assumption. Indeed, as we will see in the final part of the paper, the action of non-ITG transport mechanisms needs to be invoked to explain the observed impurity profiles.

In the following we describe the effect of impurities on ITG mode destabilization in Sec.~\ref{sec:imponitg}, then the role of ITG instabilities in determining impurity transport in Sec.~\ref{sec:itgonimp}, and finally the nonlinear mutual interaction between them in Sec.~\ref{sec:nonlinear}.


\section{Effect of impurities on ITG modes}
\label{sec:imponitg}
\begin{figure}
\includegraphics{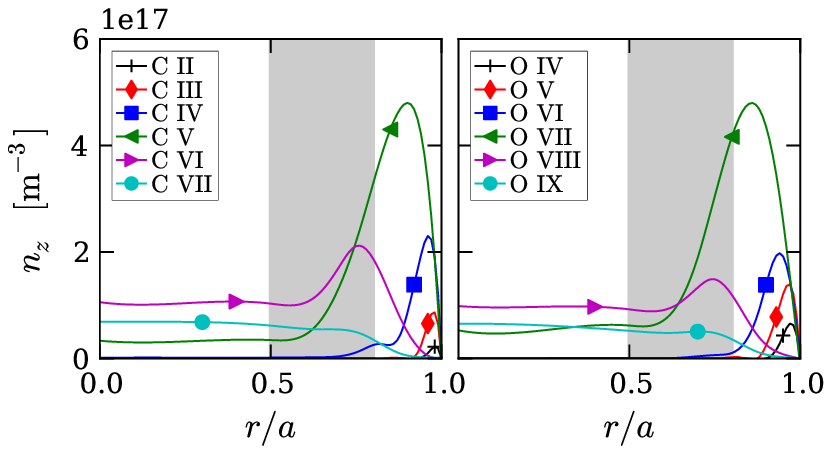}
\caption{Impurity profiles of Carbon (left) and Oxygen (right), shot 23977 at 200 ms. The total volume impurity concentration is about 2\% of the total electron content. The occurrence of ITG modes will be analyzed in the shaded radial region.}
\label{fig:impprofiles}
\end{figure}

In RFX-mod the density profiles of the most populated impurity ions are usually peaked outside the core, as shown in Fig.~\ref{fig:impprofiles}, Carbon and Oxygen being the most relevant species. Based on passive spectroscopy evaluations, their volume content is evaluated to be around a 1-2\% of the total number of electrons. The present Carbon and Oxygen profiles are reconstructed following the method described in Ref.~\cite{carraro00}. All of the available experimental data concerning emission lines, continuum radiation and total radiated power emitted by the intrinsic impurities are reproduced with a one-dimensional (radial) time-dependent collisional radiative diffusion model. The model gives the best evaluation of the impurity transport coefficients with the resulting Carbon and Oxygen ion density profiles. Such profiles are fully consistent with the output of the transport code \textsc{Ritm}~\cite{tokar94} or its MHD extension \textsc{Riport}~\cite{predebon09}. Given the electron and impurity density profiles, the ion density is derived accordingly.

In the following of this section we are going to describe the setup of the gyrokinetic simulations, the effect of a single impurity species on ITG modes, and finally the effect of multiple impurity species (CV-VII and OVII-IX) on these modes. The reference experimental cases are the well-examined shots 23977 at 200 ms and 24932 at 64 ms. Both are single helical axis cases which experience a barrier in the electron temperature profile at mid-radius. The first one has the electron density peaked in the core, while the second has a slightly hollow $n_e$ profile until $\rho=r/a\sim 2/3$. In shot 23977 (cf. Fig.~\ref{fig:impprofiles}) the most abundant Carbon ions in the outer core, CV (or C4+) and CVI (C5+), have logarithmic derivatives $a/L_{n_\mathrm{CV}}=-a(dn_{\mathrm{CV}}/dr)/n_{\mathrm{CV}}\simeq -12$ and $a/L_{n_\mathrm{CVI}}\simeq-6$ in the radial interval $0.6<r/a<0.7$; similarly the most abundant Oxygen ions, OVII (or O6+) and OVIII (O7+), have $a/L_{n_\mathrm{OVII}}\simeq-11$ and $a/L_{n_\mathrm{OVIII}}\simeq-5$ in the same region. In shot 24932 the impurity radial profiles are even steeper, their logarithmic gradients locally reaching values $\simeq-15$ for CV and OVII, in $r/a\in[0.6,0.9]$.


\subsection{Background}
Before discussing the setup and the results of the simulations, we briefly review the modifications required for GS2 to work in reversed field pinch geometry. We recall that such geometry is characterized by a $q$ profile everywhere $\ll 1$, decreasing from the core to the edge, where it becomes slightly negative. As extensively discussed in~\cite{predebon10a}, having the poloidal and toroidal component of the magnetic field the same strength, $B_\theta\sim B_\phi$, the definition of some terms in the gyrokinetic equation has been modified with respect to the standard tokamak geometry:
(i) the curvature and $\nabla B$ drifts in the drift frequency $\omega_d$ include terms of order of $\mathcal{O}(1)$ proportional to $B_\theta^2/\rho$ and $-\partial B/\partial\rho$ respectively;
(ii) the parallel gradient is $\mathbf{b}\cdot\nabla \propto (B_\theta/\rho)\partial_\theta$, where the poloidal angle $\theta$ is used as longitudinal (ballooning) coordinate along the flux tube;
(iii) the equilibrium field is given by $B=B_0(1-\delta\cos\theta)$ with $\delta<\epsilon\rho$, $\epsilon=a/R$;
(iv) the binormal component of the wavenumber is given by $k_y=(n/r)(\epsilon^2\rho^2+q^2)^{1/2}$, where $n$ is the toroidal mode number.

Coming to the description of the quantities related to particle distributions, we recall that an arbitrary number of species can be included in GS2, with a full local description of charge $z_j$, mass $m_j$, density $n_j$ and density logarithmic gradient $a/L_{n_j}=-a(d n_j/dr)/n_j$, temperature $T_j$ and temperature logarithmic gradient $a/L_{T_j}=-a(d T_j/dr)/T_j$, and collisionality $\nu_j$. Plasma quasi-neutrality has to be preserved across the flux tube:
\be
\label{eq:qn}
\sum_j z_j n_j = 0, \quad
\sum_j z_j n_j\frac{a}{L_{n_j}} = 0.
\ee

In this work most of the linear simulations are performed including a single impurity only, with charge $z$ and density $n_z$, so that the expressions above simply read $n_e = n_i + z n_z$, and $a/L_{n_e} = (n_i/n_e)a/L_{n_i} + (z\nzne)a/L_{n_z}$. The electron density and temperature are the experimental ones. Consistently with the RFX-mod experimental operation, a H background is considered, $m_i=1$. The conditions $T_e/T_i=1.4$ and $T_z=T_i$ are assumed $\forall\rho\in[0,1]$. Simulations are performed collisionless, which reduces the computational time, since the effect of electron collisions has been tested on a few representative cases and found to be small. Trapped electrons are always included. Concerning the impurity density profiles, these have to be considered as ``starting'' profiles: in the following we will in fact scan over gradient and density, so as to provide a sensitivity study beyond the experimental uncertainties.

The way we carry out single impurity calculations is to fix one impurity ion at a time (among the most relevant ionization states), and to rescale its density profile so that the volume concentration is a 2\% of the total number of electrons, $\int_0^a n_z r\,dr/\int_0^a n_e r\,dr = 2\cdot 10^{-2}$. Of course this implies an overestimate of the density of the ion under consideration, but allows to have lighter gyrokinetic simulations and to better understand the physical mechanism at work.

\begin{figure}
\includegraphics{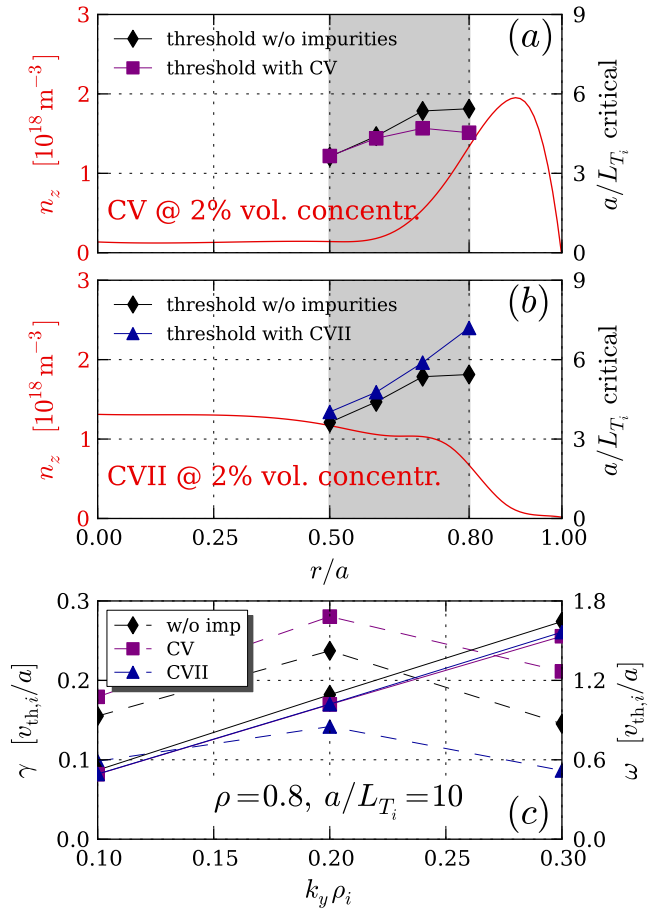}
\caption{ITG threshold study for shot 23977. In (a) and (b) only CV and CVII are considered respectively, their volume concentration being set to a 2\% of the total electron content; $\blacklozenge$ symbols represent the thresholds calculated without any impurities. In (c) real frequency (solid) and growth rate (dashed) spectra refer to the most external radius at the highest ion temperature gradient of the scan.}
\label{fig:oneimp}
\end{figure}

As a first result, in Fig.~\ref{fig:oneimp} we show the ITG stability study taking into account CV (the most populated state of Carbon) or CVII (fully stripped Carbon), for shot 23977. The stability threshold is calculated as the minimum value of the ion temperature gradient at which the growth rate of the mode is positive. The comparison with the case without impurities shows that the inclusion of an impurity species with positive radial derivative ($\alnz<0$) causes ITG mode destabilization. Conversely, a negative radial derivative ($\alnz>0$) induces ITG mode stabilization. A higher stability threshold implies an overall lower growth rate of the ITG modes, and viceversa, while the real frequency remains almost untouched, see Fig.~\ref{fig:oneimp}-c. For the present profiles the effect of the inclusion of a single impurity is not dramatic: in either cases the thresholds remain much larger (a factor $\sim\epsilon^{-1}=R/a$) than for a tokamak. Landau damping thus remains the key mechanism discriminating between the two configurations. As we will see, a larger impurity density derivative is required in order to get an important ITG mode excitation, or even the occurrence of impurity drift instabilities.


\subsection{Parametric study with a single impurity species}
To understand the interplay between impurities and ITG modes, we extend the single impurity analysis to a larger parameter domain. In particular we focus on the dependence of the mode frequency/growth rate and stability threshold on the local quantities $n_z(\rho)/n_e(\rho)$ and $\alnz(\rho)$. Based on the experimental case of Fig.~\ref{fig:oneimp}, we perform a set of linear calculations at $\rho=0.7$ for CV, and at $\rho=0.6$ for CVII. The local impurity density is varied in the interval $\nzne\in[0,3\%]$, and the gradient in the interval $\alnz\in[-20,+20]$.

\begin{figure}
\includegraphics{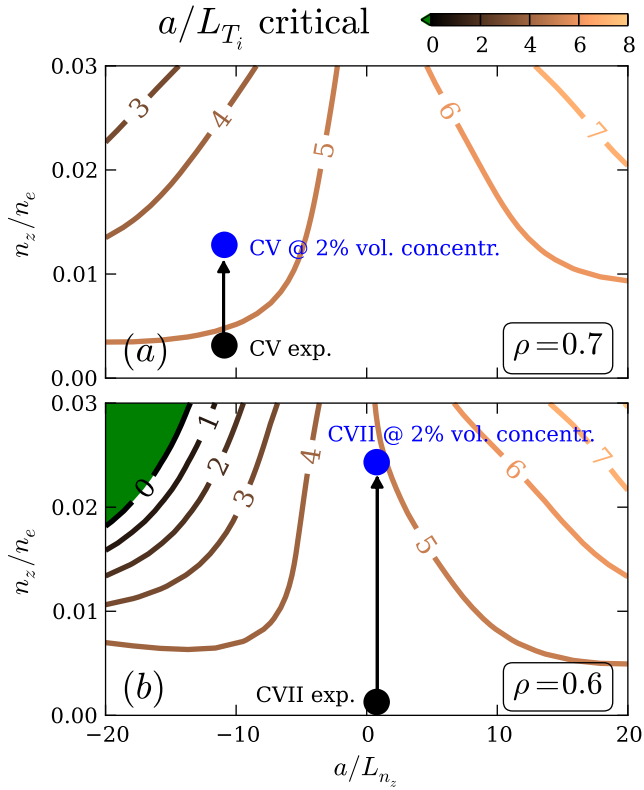}
\caption{$\altic$ contour as a function of the local values of $\nzne$ and $\alnz$ for a single impurity at fixed radius, shot 23977. At $\rho=0.7$ (a) the experimental value of CV is increased to get it covering the whole 2\% of the impurity volume concentration, then the analysis is extended to $(\nzne,\alnz)\in[0,0.03]\times[-20,20]$. At $\rho=0.6$ (b) the experimental value of CVII is increased to get it covering the whole 2\% of the impurity volume concentration, then the analysis is extended to $(\nzne,\alnz)\in[0,0.03]\times[-20,20]$. In the green area the instability is obtained with $\altic<0$.}
\label{fig:scan23977}
\end{figure}

The result of this set of computations is reported in Fig.~\ref{fig:scan23977}, where the critical ion temperature gradient is shown as a function of both the parameters of interest. Large impurity densities and positive radial gradients of impurity density profiles ($\alnz<0$) are favorable to ITG destabilization. If the whole impurity content at $\rho=0.7$ were CV (Fig.~\ref{fig:scan23977}-a, ``CV @ 2\% vol. concentr.'') or CVII at $\rho=0.6$ (Fig.~\ref{fig:scan23977}-b, ``CVII @ 2\% vol. concentr.''), the destabilization would not be so relevant. However, in the case of CVII, we can see that for high enough $\nzne$ and $\alnz\ll -1$, ITG modes are well destabilized, and mode destabilization can be found also for a vanishing ion temperature gradient. The effect is visible only in the CVII case, due to the higher charge of the ion and to the consequent higher charge fraction $z\nzne$. The resulting instability is the so-called impurity mode, well discussed in the literature~\cite{coppi66,tang80,jarmen,dong94,dong95,horton99}. The effect of the impurity profile becomes even larger for larger ion charges, e.g., for the most ionized states of Oxygen.

\begin{figure}
\includegraphics{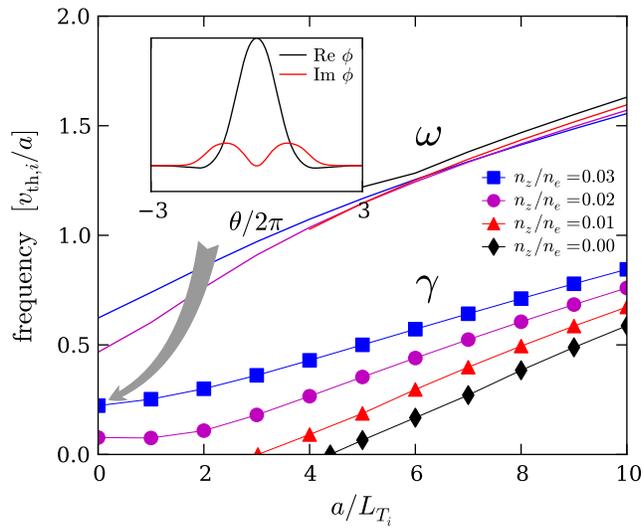}
\caption{Section $\alnz=-20$ of the contour in Fig.~\ref{fig:scan23977}-b. Increasing $\nzne$ drives destabilization also for vanishing $\alti$. In the inset the mode structure is plotted as a function of the ballooning angle at $\alti=0$.}
\label{fig:impmode}
\end{figure}

Let us investigate more quantitatively the effect of an increasing local impurity concentration at fixed outwardly peaked impurity, $\alnz=-20$, for the case CVII at $\rho=0.6$, Fig.~\ref{fig:impmode}. Increasing $\nzne$ leads to a progressive destabilization of the ITG mode, until the growth rate $\gamma(\alti)$ becomes positive for a vanishing ion temperature gradient. In the inset of Fig.~\ref{fig:impmode} the resulting impurity mode structure at $n_z/n_e=3\%$ as a function of the ballooning angle does not reveal a major modification with respect to the usual ITG mode, neither the mode frequency has discontinuities at low ion temperature gradients. As reported in Ref.~\cite{dong95} the impurity mode and the $\eta_i$ mode (i.e., the ITG mode, $\eta_j=L_{n_j}/L_{T_j}$ henceforth) are strongly coupled, making it difficult to distinguish between them.

As a general comment on the effect of the densities on ITG instabilities, for a fixed $n_e$ profile the shape of the impurity density has strict consequences on the main ion density profile and derivative. We recall that in the RFP the (hydrogenic) ITG threshold $\altic$ as a function of $\alni$ has a minimum for $\alni>0$ ($\alni\simeq 4$ in this case), mainly due to the strong $\omega_{\nabla B}$ resonance occurring around such density gradient, cf. Fig.~6-b of Ref.~\cite{predebon10a}. Therefore, given the experimental $n_e$ profile, in order to preserve quasi-neutrality, a negative $\alnz$ (outwardly peaked impurity) causes $\alni$ to increase, and such increase (unless very large) provides ITG mode destabilization. Viceversa, positive $\alnz$ (inwardly peaked impurity) causes $\alni$ to decrease, and the ITG mode to be progressively stabilized.

Looking again at Fig.~\ref{fig:scan23977}, for the experimental almost flat $n_e$ profile ($|\alne|<1$), if the impurity density is high enough, the effect of a large impurity density gradient is to cause an inversion of the sign of the main ion density gradient. In both cases, CV and CVII, the (left) half-plane $\alnz<0$ roughly corresponds to $\alni>0$, with a maximum for $\alni$ corresponding to the upper-left corner of the respective frames ($\alni\sim 3$ for CV, $\alni\sim 4$ for CVII). Conversely, in the (right) half-plane $\alnz>0$ the value of $\alni$ becomes negative ($\eta_i<0$), with the minimum in the upper right-corner. This study has been repeated for other ionized impurities, shots and electron density profiles, indicating that for a strong ITG mode destabilization or even the impurity mode onset to occur, there must be a combination of the destabilizing term $\alni>0$ and the relative sign of impurity and main ion density gradients, $\omega_{*i}\omega_{*z}<0$ (with $\omega_{*j}=k_y cT_j /eBL_{n_j}$), cf. Ref.~\cite{coppi66}. In particular the impurity mode at $\eta_i=0=\eta_z$ is produced for reversed impurity density gradients, and if the impurity charge fraction is high enough. This phenomenology confirms the former results obtained in tokamak plasmas~\cite{coppi66,tang80,jarmen,dong94,dong95}, the physical mechanisms at work being essentially the same in both cases: however, the manifest drop of the stability threshold observed in the RFP makes impurities particularly important, as they can determine the transition from a condition where ITG modes are inherently absent to a scenario possibly dominated by such turbulence.

The parameter region providing impurity-mode destabilization is not reached in the experimental cases examined so far. Such results are obtained by modifying the single impurity parameters of interest. To get a more realistic conclusion we have to consider the case of many partially-ionized impurity ions simultaneously present at a time, as in the experiment.


\subsection{Simulations with several impurity species}
The 6 most ionized ions, Carbon V to VII and Oxygen VII to IX, are included in the linear ITG simulations, overall rescaled so as to get a total 2\% concentration, $\sum_{j\ne i,e} \int_0^a n_j r\,dr/\int_0^a n_e r\,dr = 2\cdot 10^{-2}$. Again, for any impurity species the collisionality is neglected, $\nu_j=0$, and the respective temperature profile is assumed equal to the ion one, $T_j=T_i$, $\altj=\alti$. The quasi-neutrality across the flux tube is imposed to the main ion density, making Eq.\ref{eq:qn} satisfied all along the radius.

We would expect a cumulative effect of such impurities if their profiles had all either a positive or a negative slope, but this is not generally the case. The single impurity profiles are not peaked at the same radius, and the stripped CVII and OIX profiles are actually rather flat. This is for example the case for shot 23977, and also for the other shot we are going to analyze now, 24932 at 64 ms.

\begin{figure}
\includegraphics{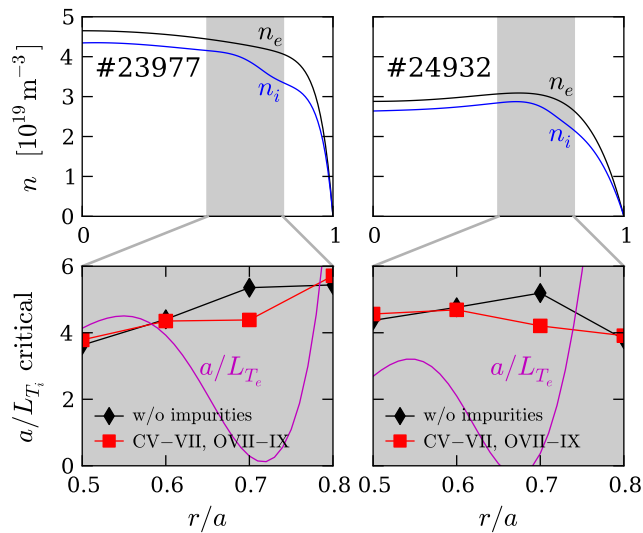}
\caption{Electron density for shots 23977 and 24932, and respective radial dependence of the ITG threshold without impurities ($\blacklozenge$ symbols), and in the presence of the 6 most ionized impurities, CV$\to$VII and OVII$\to$IX ($\blacksquare$ symbols). The experimental profile of $\alte$ is plotted for comparison.}
\label{fig:manyimp}
\end{figure}

In Fig.~\ref{fig:manyimp} the two shots are compared. They are characterized by different electron density profiles, the latter being slightly hollow. The net effect of the impurities is in both cases a somewhat ``destructive interference''. No dramatic ITG destabilization is found in the region of interest, neither in the peaked (23977) nor in the hollow (24932) electron density profile case; only a slight destabilization is present at $\rho\sim 0.7$, where $\alni\gtrsim 1$.

\begin{figure}
\includegraphics{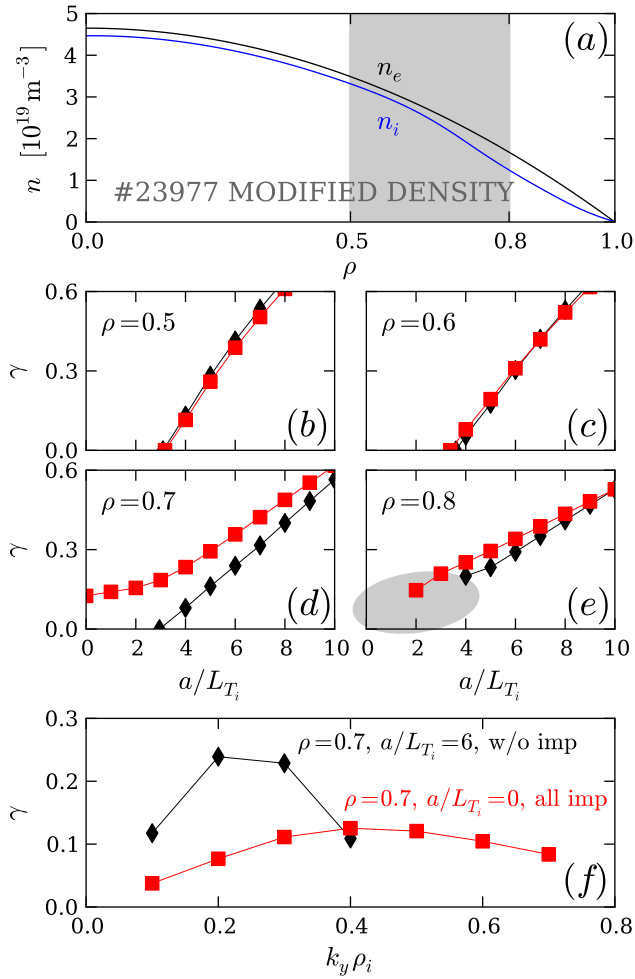}
\caption{Modified peaked electron profile for shot 23977 (a), with the maximum growth rate $\gamma$ as a function of $\alti$ from $\rho=0.5$ to $\rho=0.8$ (b-e) for the case without impurities ($\blacklozenge$ symbols), and with the most stripped impurities, CV$\to$VII and OVII$\to$IX ($\blacksquare$ symbols). The shaded area in (e) represents the range in $\alti$ where modes do not propagate in the ion diamagnetic direction. Two spectra at $\rho=0.7$ are in frame f, for $\alti=6$ without impurities ($\blacklozenge$) and for $\alti=0$ with all the impurities ($\blacksquare$)}
\label{fig:peaked}
\end{figure}

We have performed another set of linear gyrokinetic simulations based on a different electron density profile. Starting from the data of shot 23977, and in particular keeping the same impurity profiles, we have built a peaked density, as shown in Fig.~\ref{fig:peaked}-a. Impurity profiles are rescaled on the new $n_e$ profile to get a total 2\% concentration. A similar electron density is not strictly experimental, peaked profiles emerging only in very low density discharges, but remains a useful test-bed for our purpose. Following exactly the same procedure carried out before, we look at the different characteristics of ITG modes without impurities and with 6 impurities (CV$\to$VII and OVII$\to$IX). While the introduction of impurities does not provide any significant changes in the core until $\rho\sim 0.6$ (Fig.~\ref{fig:peaked}-b/c), in this case their superposition causes an important mode destabilization at $\rho=0.7$, where $\alni\simeq 4$ and $\omega_{*i}\omega_{*z}<0$ for the most abundant impurities. Such a destabilization is obtained also for vanishing $\eta_i$, see Fig.~\ref{fig:peaked}-d. The spectrum of the resulting impurity mode at $\alti=0$ is plotted in frame f, and covers a larger $\aky$-range with respect to the usual ITG mode excited, for example, at $\alti=6$. At the extreme of the radial interval, $\rho=0.8$ (Fig.~\ref{fig:peaked}-e), the dependence of the growth rate as a function of $\alti$ appears less simple: the values of $\gamma$ for low $\alti$ are missing because we are not considering here instabilities with negative real frequency, i.e., propagating in the $\omega_{*e}$ direction. In the corresponding range of $\alti$ (covered by the shaded area in the figure), trapped electron modes appear to be the fastest growing instabilities, indeed excited by the high value of the electron density gradient at that radius. The occurrence of trapped electron modes in the reversed field pinch is a topic to be extensively investigated in the future.

Summarizing this first part of the paper, no dramatic influence of the impurities is found on ITG mode destabilization for rather flat electron density profiles. In the next section we will discuss a method to indirectly evaluate the nature of the microturbulence at work, based on the direction of the impurity flux. This could be considered as a counter-evidence of the impurity-driven nature of microinstabilities.


\section{Effect of ITG modes on impurities}
\label{sec:itgonimp}
A completely different approach is used to evaluate linearly the direction of the impurity flux in ITG dominated discharges. While for the previous case the impurity concentration had to be large enough to substantially perturb the system, for this study impurities are assumed to have negligible charge fraction, $z\nzne\ll 1$, so as not to influence the turbulence itself. In other words, we are supposed to deal with \emph{trace} impurities.


\subsection{Background}
We summarize how, from linear gyrokinetic studies, it is possible to deduce the so-called peaking factor of an impurity species. The procedure is more extensively described, for example, in Refs.~\cite{angioniprl06}, \cite{angionippcf07}, and \cite{hein10}.

For any species $j$, the quasi-linear particle flux $\Gamma_j=\langle \tilde n_j \tilde v_{E\times B}\rangle$, with $\langle\cdot\rangle$ magnetic flux surface average, can be written in the form
\be
\frac{a\Gamma_j}{n_j} = D_j\frac{a}{L_{n_j}} + D_{Tj}\frac{a}{L_{T_j}} + av_{Pj}
\ee
with $D_j$ diagonal diffusion coefficient, $D_{Tj}$ off-diagonal thermodiffusion coefficient, and $v_{Pj}$ pure convection. In general, these coefficients do depend on $\alnj$ and $\altj$ through the dependence of the eigenvalues on such gradients. In the case of a single impurity with charge $z$, if its charge concentration is so small ($z\nzne\ll 1$) that it can be neglected in the gyrokinetic Poisson/Ampere equations, the coefficients $D_z$, $D_{Tz}$ and $v_{Pz}$ become independent of $\alnz$ and $\altz$. Thus the flux of a trace impurity species simply becomes a linear function of the gradients.

The equation above for the impurity flux can be written as
\be
\frac{a\Gamma_z}{n_z} = D_z\left(\frac{a}{L_{n_z}}+C_T\frac{a}{L_{T_z}}+C_P\right)
\ee
where we have defined $C_T=D_{Tz}/D_z$ and $C_P=av_{Pz}/D_z$. Under the mentioned assumption of trace concentration, such coefficients can be derived from the slopes of the quasi-linear flux $\Gamma_z$ (calculated by GS2) as a function of the logarithmic gradients $\alnz$ and $\altz$, and from the residual flux at vanishing gradients. The way to get them is to turn on/off the gradients, always preserving quasi-neutrality. The total convective contribution is finally given by the sum of the thermodiffusive term and the pure convection,
\be
\frac{av_z}{D_z} = C_T\frac{a}{L_{T_z}} + C_P.
\ee
In stationary conditions, and without particle sources, diffusion and convection must balance so as to yield $\Gamma_z=0$. This reflects on a constraint for the impurity density gradient, $\alnz=-C_T\,\altz-C_P=-av_z/D_z$. The sign of the coefficients $C_T$ and $C_P$ and that of the peaking factor $-av_z/D_z$ are strongly affected by the dominant instability, and can be considered the signature of the related turbulence in the plasma.


\subsection{Impurity peaking in ITG dominated plasmas}
Following the procedure just described, we now study how the impurity flux is affected by ITG instabilities. In the absence of plasma rotation ITG modes are commonly believed to yield an inward impurity flux, at least in tokamaks~\cite{puiattivalisa06, angionipop07, angioninf09, valisa11}. The RFP turns out not to be an exception to this rule.

For our set of linear simulations the charge density is set to $z\nzne=10^{-4}$, so as not to significantly perturb the system. The ITG instabilities are artificially excited by imposing an ion temperature gradient above the threshold at every radius in the region of interest, in this case $\alti=2\alte$. This is not consistent with any $T_i$ profile, but is certainly allowed in the framework of flux tube simulations.

\begin{figure}
\includegraphics{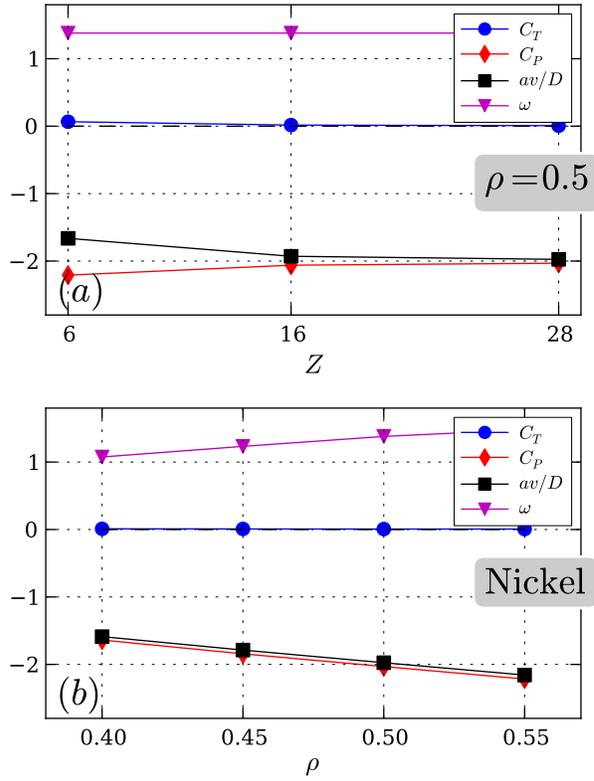}
\caption{Impurity peaking $av/D$ with respective $C_P$, $C_T$ and ITG mode frequency $\omega$ for different values of impurity charge at fixed radial position $\rho=0.5$ (a), and for different radial positions at fixed charge $Z=28$ (b). The unphysical condition $L_{T_e\mathrm{exp}}/L_{T_i}=2$ is imposed to the ion temperature gradient at each radius.}
\label{fig:impinward}
\end{figure}

The dependence of $C_T$, $C_P$, and $av_z/D_z$ on the trace impurity charge $Z$, at fixed radius, is shown in Fig.~\ref{fig:impinward}-a. The impurity charge is varied between 6 and 28, i.e., between Carbon and Nickel. The frequency is positive and constant. The mode structure itself is not changed, indicating that the ITG mode is not modified by the inclusion of a low-density impurity. The thermodiffusion is positive and decreases with $Z$, with an opposite sign with respect to the curvature pinch $C_P$. The quantity $av_z/D_z$ turns out to be negative, i.e. an inward pinch is driven; analogously, the peaking factor is positive, resulting in an impurity accumulation in the core and a peaked impurity profile in stationary source-free conditions. This is not compatible with the experimental results discussed in Ref.~\cite{menmuir10}, which report a strong outward impurity convection, especially in the outer plasma core.

The impurity convective contribution $av_z/D_z$ is also evaluated across the ITB, at several radii for a fixed charge, e.g., Nickel, see Fig.~\ref{fig:impinward}-b. Until the dominant instability is ITG, the inward direction of the pinch cannot be prevented. Of course, the resulting radial trend of $av_z/D_z$ has to be merely seen as a qualitative indication, being strictly dependent on the selected local ion temperature gradient, which is actually unknown; however, the sign of the peaking factor is well-established. Assuming microturbulence to be the dominant transport mechanism, other instabilities are likely to be at work.


\section{Nonlinear studies of ITG turbulence with impurities}
\label{sec:nonlinear}
Although linear gyrokinetic calculations are essential to determine the impact of impurities on ITG mode destabilization, and helpful to derive quasi-linearly the basic particle transport properties of a trace species, the problem of the mutual interaction between ITG modes and impurities should be rigorously treated in a self-consistent nonlinear framework. In the final part of this paper we start addressing the topic of ITG turbulence in 3-species RFP plasmas, reconsidering two cases discussed in Sect.~\ref{sec:itgonimp} and \ref{sec:imponitg} respectively.

\begin{figure}
\includegraphics{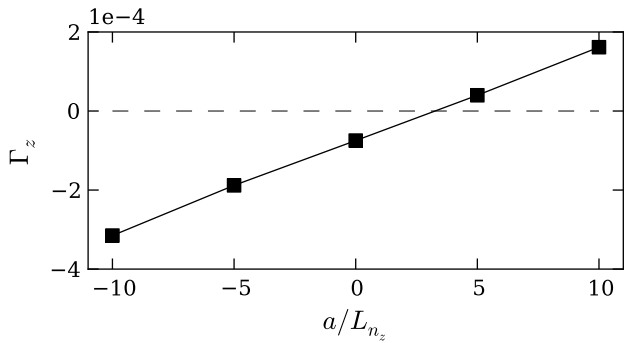}
\caption{Nickel flux as a function of its density gradient (with $z\nzne=10^{-4}$, $z=28$) at $\rho=0.5$, from three-species nonlinear simulations of ITG turbulence assuming $L_{T_e\mathrm{exp}}/L_{T_i}=2$. The impurity flux is linear in $\alnz$, and $av_z/D_z\simeq-3$.}
\label{fig:nonlinear2}
\end{figure}

Previously we have claimed that trace-impurity fluxes are linear in density/temperature gradients. As a numerical proof, we have performed a set of nonlinear simulations for Nickel at $\rho=0.5$, with exactly the same parameters used for Fig.~\ref{fig:impinward}. Only the impurity density gradient is varied in this case. Thermoconductivity is therefore included in the convective term, with the impurity temperature gradient set to $\altz=\alti\simeq 8.3$. At $\rho=0.5$ the magnetic shear is $\hat s=-0.64$. The wavenumber grid has $\Delta k_x\rho_i\simeq\Delta k_y\rho_i=0.1$, $\max(k_y\rho_i) = 0.6$, with a box size $L_y\simeq 64\,\rho_i$, $L_x/L_y\simeq 1$. The peak of the linear spectrum is at $\aky\sim 0.3$. The result of the nonlinear calculations is shown in Fig.~\ref{fig:nonlinear2}, where the perfect linearity of the flux in $\alnz$ yields $av_z/D_z\simeq-3$, i.e. a larger value (in absolute value) than the linear results obtained in the previous section, but not very far from them.

It is widely more interesting to come back to the single-impurity simulations with large $\nzne$ of Sec.~\ref{sec:imponitg}, as an integration to the previous linear analysis. To this aim, we deal again with shot 23977, Carbon VII at $\rho=0.6$ with $\alti=10$, and in particular for $\nzne=0.03$ with impurity gradients in the interval $\alnz\in[-10,+10]$, see Fig.~\ref{fig:scan23977}-b for a comparison with the linear cases. The electron density gradient is $\alne=0.23$.

At $\nzne=0.03$ impurities do perturb the system, the ITG stability threshold, and the possible onset of impurity drift modes. Linear $k_y$ spectra are different in the considered range of $\alnz$, being more extended for $\alnz=-10$ (until $\aky=0.7$) than for $\alnz=10$ (until $\aky=0.5$). For the nonlinear simulations the wavenumber grid has $\Delta k_x\rho_i\simeq\Delta k_y\rho_i=0.075$, with $\max(k_y\rho_i)\simeq 1$ in the case $\alnz=-10$, and with a box size $L_y\simeq 80\,\rho_i$, $L_x/L_y\simeq 1$. At this radius the magnetic shear is $\hat{s}=-1.22$.

\begin{figure}
\includegraphics{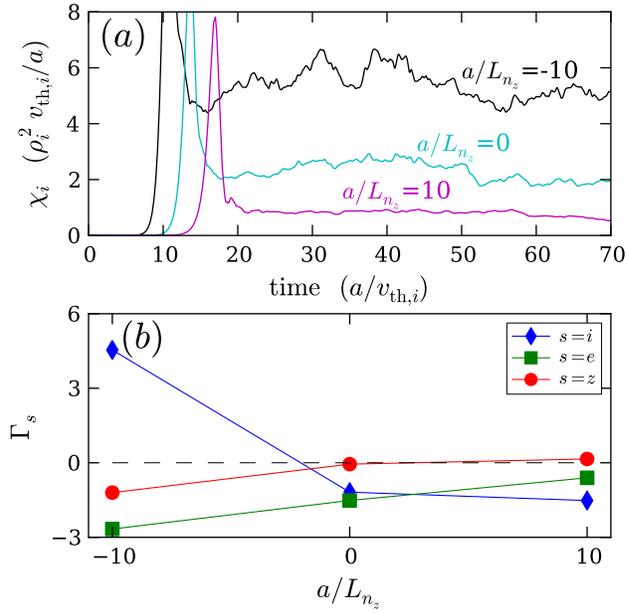}
\caption{Three-species nonlinear simulations of ITG turbulence with CVII having $\nzne=0.03$, $\alnz\in[-10,10]$ (cf. Fig.~\ref{fig:scan23977}-b), and with ion temperature gradient $\alti=10$. The ITG destabilization for negative $\alnz$ corresponds to an increase of the ion thermal conductivity (a). Outwardly peaked impurity profiles ($\alnz<0$) drive outward fluxes of main ions; inwardly peaked impurity profiles ($\alnz>0$) reinforce the inward fluxes of main ions (b).}
\label{fig:nonlinear1}
\end{figure}

In Fig.~\ref{fig:nonlinear1}-a we report the temporal evolution of the ion conductivity $\chi_i=-Q_i/(n_i\,\partial T_i/\partial r)$, where $Q_i$ is the ion heat flux, in gyro-Bohm units, $\chiGB=\rho_i^2\vthi/a$. We clearly see the expected decrease of $\chi_i$ for increasing $\alnz$, in analogy with the increase of the stability threshold. In Fig.~\ref{fig:nonlinear1}-b the saturated particle fluxes are reported as a function of $\alnz$. We notice that: (i) outwardly peaked impurities ($\alnz<0$), with inwardly peaked main ions due to quasi-neutrality ($\alni>0$), are found to have a net inward flux ($\Gamma_z<0$), and to provide an outward ion flux ($\Gamma_i>0$); (ii) for flat impurities ($\alnz=0$) the impurity flux is inward ($\Gamma_z\simeq -0.06$) as well as the main ion flux ($\Gamma_i<0$); (iii) inwardly peaked impurities ($\alnz>0$), with outwardly peaked main ions due to quasi-neutrality ($\alni<0$), have an outward flux ($\Gamma_z\simeq 0.15$), and cause ions to be pinched ($\Gamma_i<0)$. In all the cases electrons are pinched, $\Gamma_e<0$. The anomalously large value $|\Gamma_z|$ obtained for the reversed Carbon gradient ($\alnz=-10$) can be interpreted as the result of the impurity mode branch working in the high wavenumber range, which indeed has the consequence of producing further inward convection of Carbon and an outward flux of Hydrogen \cite{tang80}.

As for the previous linear simulations, this rather strong effect is associated with the high impurity charge concentration and to the fact that the whole impurity content is moved onto a single ionized species, C VII in this case, with steep density profile.


\section{Conclusions}
The presence of the impurities has been considered in the framework of linear and nonlinear gyrokinetic simulations, applied to the RFP configuration. ITG modes, whose stability threshold $\altic$ in a two-component plasma is a factor $\sim\epsilon^{-1}$ larger than in the tokamak, turn out to be slightly destabilized in the presence of outwardly peaked impurity profiles. This behaviour is in agreement with the past tokamak and RFP results. Parametric studies show that a more serious destabilization, including the appearance of impurity drift modes, could be obtained for very peaked main gas densities, a condition not commonly observed in correspondence to the positive impurity slopes of the outer core plasma.

On the other hand, if ITG instabilities/turbulence only were active, the convection of impurities should occur towards the plasma core. Instead, experimentally it is found to be outward. This suggests that other instabilities may be active; further work is required to understand their existence and their role on impurity flux reversal. In particular the presence of magnetic-surface breaking instabilities has to be considered; even in single helicity states magnetic field stochasticity could drive a significant fraction of the impurity transport in the region investigated in this paper, as well as of the main gas particle and energy transport.


\ack
The authors thank S.~C.~Guo, M.~E.~Puiatti and M.~Valisa for fruitful discussions, F.~Auriemma and A.~Alfier for providing the experimental data used in the paper, and W.~Dorland and M.~Kotschenreuther for making the code GS2 available. This work has been supported by the European Communities under the contract of Association between Euratom/ENEA.


\section*{References}
\bibliography{bibdb,rfpdb}


\newpage

\end{document}